# Long-distance thermal temporal ghost imaging over optical fibers

Xin Yao,[1] Wei Zhang,[1*] Hao Li,[2] Lixing You,[2] Zhen Wang,[2] Yidong Huang[1]

[1]*Tsinghua National Laboratory for Information Science and Technology, Department of Electronic Engineering, Tsinghua University, Beijing, 100084, China*
[2]*State Key Laboratory of Functional Materials for Informatics, Shanghai Institute of Microsystem and Information Technology, Chinese Academy of Sciences, Shanghai 200050, China*
*\*Corresponding author: zwei@tsinghua.edu.cn*

A thermal ghost imaging scheme between two distant parties is proposed and experimentally demonstrated over long-distance optical fibers. In the scheme, the weak thermal light is split into two paths. Photons in one path are spatially diffused according to their frequencies by a spatial dispersion component, then illuminate the object and record its spatial transmission information. Photons in the other path are temporally diffused by a temporal dispersion component. By the coincidence measurement between photons of two paths, the object can be imaged in a way of ghost imaging, based on the frequency correlation between photons in the two paths. In the experiment, the weak thermal light source is prepared by the spontaneous four-wave mixing in a silicon waveguide. The temporal dispersion is introduced by single mode fibers of 50 km, which also could be looked as a fiber link. Experimental results show that this scheme can be realized over long-distance optical fibers.

*OCIS codes: (190.4380) Nonlinear optics, four-wave mixing; (110.1650) Coherence imaging.*

Ghost imaging (GI) is an interesting imaging technique. It images objects by the correlation of two light beams [1]. The first GI was realized by spatially correlated photon pairs generated by parametric down-conversion (PDC) [2]. Since then, GI has attracted intensive attention, from its physical principles and properties to its realizations and potential applications [3-7]. An important progress emerged when the pseudo-thermal light source replaced the quantum counterpart in spatial GI experiments, which enriched the understanding of two-photon or two-beam correlation and reduced the requirement of the sources for GI [4,8,9]. More importantly, it demonstrated that GI can be realized by not only quantum correlation but also classical correlation. It promoted the proposals and demonstrations of new GI concepts [10-13], such as computational GI [12] which removes the high spatial resolution detector for the reference beam, and differential GI [13] which enhances the signal-to-noise ratio of the image. Recently, Ryczkowski et al promoted GI from space domain to time domain with an all-fiber setup [14], in which the imaging object is the temporal signal. This scheme shows great compatibility with optical fiber networks. On the other hand, in our previous work we proposed a GI scheme based on frequency-correlated photon pairs generated by the spontaneous four wave mixing in a silicon waveguide [15]. By introducing the spatial and temporal dispersions on the two paths of GI, the frequency correlation in a photon pair transfers to the correlation of position-arrival time of the two photons. By this correlation, one dimensional GI can be realized by the time-resolved coincidence measurement of the photon pairs. Hence, it is called as temporal GI. Since the temporal dispersion can be realized by optical fibers for photon transmission, long-distance GI over optical fibers can be realized by this scheme.

In this letter, we propose that this temporal GI can also be realized in a classical way utilizing the thermal light. The sketch of the thermal temporal GI scheme is shown in Fig. 1. The light from the thermal source is attenuated to single-photon level, then split into two paths by the beam splitter. In one path (Alice), the photons are diffused into a line according to their frequencies by a spatial dispersion component, such as a grating or a prism, then illuminate the object. The transmitted (or reflected) photons are collected by a single photon detector. Since the detector does not discriminate the frequencies or the positions of these photons, the transmission (or reflection) information of the object carried by these photons cannot be recovered by the detection results of this path only. Actually, by the spatial dispersion effect, the spatial information of object is modulated on the spectrum of thermal photons. Hence, in this work we use a narrowband tunable filter to simulate the spectral modulation process of the object. In the other path (Bob), the photons are diffused in time domain by a temporal dispersion component, then detected by another single photon detector. These photons do not carry the information of the object. However, the information could be retrieved by the time-resolved coincidence measurement of the two paths. The temporal dispersion can also be provided by the optical fibers for long-distance photon transmission. Hence, this scheme also supports the imaging process between two distant parties over the fiber link. Comparing with our previous work [15], the quantum light source is replaced by the thermal light source, which is easier to be implemented and robust against the transmission loss.

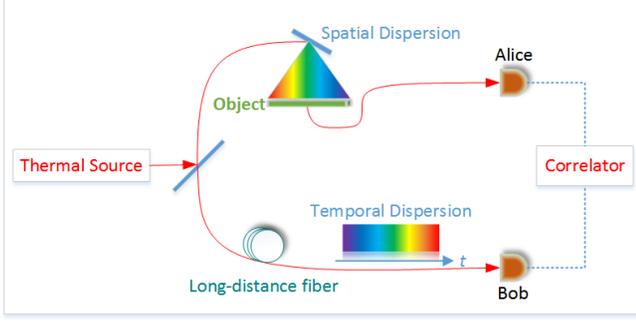

Fig. 1 The sketch of thermal temporal GI. By the correlation measurement of the thermal photons experiencing the spatial dispersion and the temporal dispersion, Bob can retrieve the image of the object at Alice side.

Firstly, we demonstrate the feasibility of the scheme theoretically by calculating the temporal second-order coherence $G^{(2)}$ of the photons in the two paths, which is proportional to the time-resolved coincident counts. The thermal light is treated as the incoherent superposition of a large number of sub-fields [16]

$$E(r,t) = \sum_j E_j(r,t) \approx e^{-i[\omega_0 t - k(\omega_0)r]} \sum_j F_\tau \left\{ f(\Omega) e^{i\varphi_j(\Omega)} \right\}, \quad (1)$$

where $j$ labels the sub-field. $\omega_0$ is the center-frequency of the thermal light and $\Omega$ is the frequency detuning of the sub-field. $F_\tau\{\ \}$ denotes the inverse Fourier transformation. Here, the first-order approximation of the dispersion is taken as $k(\omega) \approx k(\omega_0) + \frac{dk}{d\omega}\Big|_{\omega_0} \Omega$. $\tau \equiv t - \frac{r}{v_g(\omega_0)}$, where $v_g$ is the group velocity of the light and $v_g = d\omega/dk$. $f(\Omega)$ is the real-positive spectrum amplitude of the thermal light. The phase $\varphi_j(\Omega)$ varies independently and randomly with time, indicating the property of incoherent superposition of chaotic sub-fields.

At Alice side, since the spatial dispersion component directs the photons to different positions of the object along an illuminating line, the object acts as an optical filter on the photons. The profile of the filter spectrum is determined by the transmission (or reflection) pattern of the object, which is denoted as $a(\Omega)$. After transmitting through the object, the field of the photons collected by the single photon detector can be expressed by

$$E(L_a,t_a) \approx e^{-i[\omega_0 t - \beta(\omega_0)L_a]} \sum_j F_{\tau_a} \left\{ f(\Omega) a(\Omega) e^{i\varphi_j(\Omega)} \right\}, \quad (2)$$

where $L_a$ is the propagated distance from the beam splitter to the single photon detector. $t_a$ is the arriving time of the photon at the single photon detector, $\tau_a \equiv t_a - \frac{L_a}{v_g}$.

At Bob side, considering the temporal dispersion provided by the transmission fibers, the field at the single photon detector can be expressed as [17]

$$E(L_b,t_a) \approx e^{-i[\omega_0 t - \beta(\omega_0)L_b]} \sum_k F_{\tau_b} \left\{ f_k(\Omega) e^{-i\beta_2 \Omega^2 L_b/2} e^{i\varphi_k(\Omega)} \right\}, \quad (3)$$

where $\beta_2$ is the group-velocity dispersion (GVD) parameter of optical fibers and the phase term of $-i\beta_2 \Omega^2 L_b/2$ introduced by the GVD modifies the temporal shape of the thermal light. $L_b$ is the propagated distance from the beam splitter to the Bob's detector, $t_b$ is the arriving time of the photon at the detector and $\tau_b \equiv t_b - \frac{L_b}{v_g}$.

The second-order temporal correlation between the fields of the two paths should be calculated by $G^{(2)}(L_a,t_a,L_b,t_b) = \langle I(t_a,L_a) I(t_b,L_b) \rangle$. According to the randomness of the phase $\varphi_j$, it is easily obtained that

$$G^{(2)}(L_a,t_a,L_b,t_b) = \langle I(t_a,L_a) \rangle \langle I(t_b,L_b) \rangle + |\Gamma(L_a,t_a,L_b,t_b)|^2, \quad (4)$$

here, the mutual-coherence function can be calculated as [16]:

$$\Gamma(L_a,t_a,L_b,t_b) = \langle E(L_a,t_a) E^*(L_b,t_b) \rangle$$
$$= \sum_j E_j(L_a,t_a) E_j^*(L_b,t_b)$$
$$\propto \sum_j F_{\tau_a - t_{0j}} \{ f(\Omega) a(\Omega) \} F_{\tau_b - t_{0j}}^* \{ f(\Omega) e^{-i\beta_2 \Omega^2 L_b/2} \}$$
$$\propto \int_\infty dt_0 F_{\tau_a - t_0} \{ f(\Omega) a(\Omega) \} F_{\tau_b - t_0}^* \{ f(\Omega) e^{-i\beta_2 \Omega^2 L_b/2} \}$$
$$\propto F_\tau \{ f^2(\Omega) a(\Omega) e^{i\beta_2 \Omega^2 L_b/2} \}$$
$$\propto F_\tau \{ f^2(\Omega) a(\Omega) \} \otimes F_\tau \{ e^{i\beta_2 \Omega^2 L_b/2} \}, \quad (5)$$

where $t_{0j}$ is the emission time of the wavepacket from the jth sub-source, therefore the summation for a large number of sub-fields in the third step becomes an integral over $t_0$; $\otimes$ denotes the convolution and $\tau = (t_a - t_b) + (\frac{L_b}{v_g} - \frac{L_a}{v_g})$.

With a well-known identity [18]

$$\int_{-\infty}^{+\infty} \exp(-ax^2 - bx) dx = \sqrt{\frac{\pi}{a}} \exp(\frac{b^2}{4a}), \quad (6)$$

the second Fourier transformation in Eq. (5) is given by:

$$F_\tau \{ e^{i\beta_2 \Omega^2 L_b/2} \} = \frac{1}{2\pi} \int e^{i\beta_2 \Omega^2 L_b/2 + i\Omega \tau} d\Omega = \frac{1}{\sqrt{-2\pi i \beta_2 L_b}} e^{i\tau^2/2\beta_2 L_b}, \quad (7)$$

Inserting Eq. (7) into (5) and calculating the convolution, we have:

$$\Gamma(\tau) \propto \int d\tau' \left( \int d\Omega f^2(\Omega) a(\Omega) e^{i\Omega \tau'} \right) e^{i(\tau-\tau')^2/2\beta_2 L_b}$$
$$\propto \int d\tau' \left( \int d\Omega f^2(\Omega) a(\Omega) e^{i\Omega \tau'} \right) e^{-i\tau \tau'/\beta_2 L_b} \quad (8)$$
$$\propto f^2(\tau/\beta_2 L_b) a(\tau/\beta_2 L_b),$$

where in the second step high-order term $e^{i(\tau')^2/2\beta_2 L_b}$ is neglected since we have assumed that a large dispersion has been introduced at Bob side.

According to Eq. (4) and (8), it can be seen that the second-order temporal correlation has two contributions. One is $\langle I(t_a,L_a) \rangle \langle I(t_b,L_b) \rangle$, which provides a background of coincident counts, and the other one is $\Gamma$, which contains the information of the object, i.e., $a(\Omega)$. When $a(\Omega)$ shifts by $\Delta\Omega$, $\Gamma(\tau)$ will move by $\Delta\tau$ in the time domain:

$$\Delta\tau \approx \Delta\Omega \beta_2 L_b, \quad (9)$$

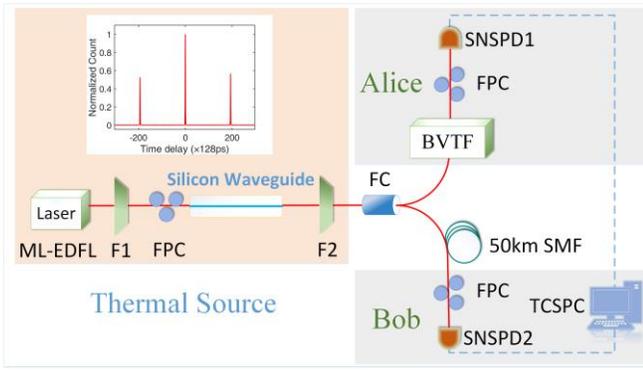

**Fig. 2.** Experimental setup. ML-EDFL, mode-locked erbium doped fiber laser; F1-2, optical filters; FC, 50:50 fiber coupler; BVTF, bandwidth-variable tunable filter to simulate the effect of the spatial dispersion component and the object; FPC, fiber polarization controller; SNSPD1-2, superconducting nanowire single-photon detectors; SMF, single-mode fiber; TCSPC, time-correlated single-photon counting module. The inset shows a HBT measurement of the thermal source and $g^{(2)}(0)=1.80$.

The principle of the thermal temporal GI is demonstrated experimentally by the setup shown in Fig. 2. It's well known that both signal and idler photons generated by spontaneous parametric down conversion (SPDC) [19] or spontaneous four-wave mixing (SFWM) [20] are under thermal statistics, and the degree of second-order temporal coherence of a thermal state is $g^{(2)}(0)=1+1/K=2$, where K is the Schmidt mode number [21]. In this experiment, the broadband fluorescence generated by SFWM in a piece of silicon waveguide is used as the thermal light. The pump light of the source is generated by a mode-lock fiber laser with a repetitive rate of 40 MHz and an average power of 4.78 dBm. It is injected into the silicon waveguide after an optical filter (F1). The center-wavelength and linewidth of the pump light is 1551.02 nm and 3.5 nm, respectively. The insertion loss of the waveguide sample is ~14 dB. The signal photons generated by SFWM in the waveguide are filtered out by another optical filter (F2) with a center-wavelength of 1530 nm. To generate photons under single-mode thermal state with K=1, the 3 dB bandwidth of F2 is about 3 nm, close to the linewidth of the pump light [22]. The property of the thermal photon source is indicated by a Hanbury Brown–Twiss (HBT) measurement, showing that the degree of second-order temporal coherence of the generated photons is $g^{(2)}(0)=1.80$, as shown in the inset of Fig. 2. Hence, the effective Schmidt mode number of the generated thermal photons is K=1.25. The mode number is a little higher than 1, which may be due to the Raman scattering in the fiber and the mismatch of filters F1 and F2. The generated thermal photons are directed to Alice and Bob through a 50:50 fiber coupler. At Alice side, the spectral modulation of the spatial dispersion component and the object is simulated by a bandwidth-variable tunable filter (BVTF, BVF300-CL, Alnair Labs Corporation), which has a variable bandwidth in the range of 0.03 nm ~ 3 nm and a tunable central wavelength. Then, the photons are sent to a superconducting nanowire single photon detector (SNSPD1). A fiber polarization controller (FPC) is placed before the SNSPD to optimize its efficiency. The other part of the thermal photons are sent to Bob over single mode fibers (SMF) of 50 km and detected by SNSPD2. The transmission times of the photons are different according to their frequencies due to the large dispersion introduced by the long-distance SMF. The system detection efficiencies of the two SNSPDs are ~40% with dark count rates of ~100 Hz and time jitters of ~80 ps. The coincidence measurements of the single photon events are realized by a time correlated single photon counting (TCSPC) module (Hydra Harp 400, Pico Quant).

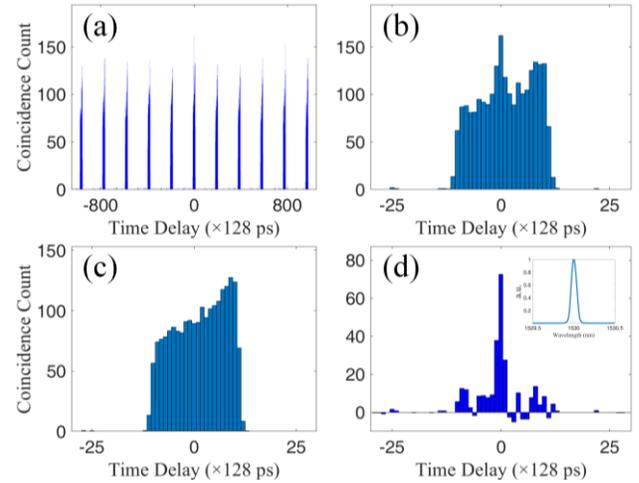

**Fig. 3.** Experiment results when BVTF's center-wavelength is set at 1530.0 nm and the bandwidth is set as 0.1 nm. (a) Coincidence histogram; (b) the coincidence peak; (c) the average of accidental coincidence peaks; (d) the difference between (b) and (c), showing the profile of $|\Gamma(\tau)|^2$, and the inset shows the BVTF's transmission spectrum.

A typical measurement result of thermal temporal GI is shown in Fig. 3, in which the center-wavelength of the BVTF is set at 1530.0 nm and its bandwidth is set at 0.1 nm. The recording time is 1200 seconds with the counting rates of ~4 KHz and ~18 KHz at Alice and Bob sides, respectively. Fig. 3(a) is the histogram of coincidence counts. There are coincidence count peaks on it, and each of them is broadened due to the temporal dispersion of the 50 km SMF. Since the repetitive rate of the pump light for the thermal photon source is 40 MHz, the intervals between the peaks are 25 ns. The peak at the center is the coincidence peak and other peaks are the accidental coincidence peaks, in which the two photons producing a coincidence event are generated by different pump pulses. It can be seen that the coincidence peak is higher than accidental coincidence peaks due to the property of thermal light. The detail of the coincidence peak is shown in Fig. 3(b) and its profile is in proportion to $G^{(2)}(L_a,t_a,L_b,t_b)$ in Eq. (4). The accidental coincidence peaks are averaged (using ten neighboring accidental coincidence peaks) and shown in Fig. 3(c), which is in proportion to the background term of $\langle I(t_a,L_a)\rangle\langle I(t_b,L_b)\rangle$ in Eq. (4). Hence, the difference between the coincidence peak and the average accidental coincidence peak would show the profile of mutual coherence $|\Gamma(\tau)|^2$ in Eq. (4), which has the information of the object according to Eq. (8). In the experiment, the information of the object is simulated by the BVTF. It spectrum is measured by an amplified spontaneous emission (ASE) source and an optical spectrum analyzer (OSA) and shown in the inset of Fig. 3 (d). Its bandwidth is 0.1 nm, far smaller than that of the thermal photon source (~3 nm) and its center-wavelength is close to that of the thermal photon source. Hence, a narrow peak at the center of the region of the coincidence peak could be expected. Fig. 3(d) shows the difference of the coincidence peak and the average accidental coincidence peak, agreeing well with the above analysis. The narrow peak is the image of BVTF's transmission spectrum by the thermal temporal GI.

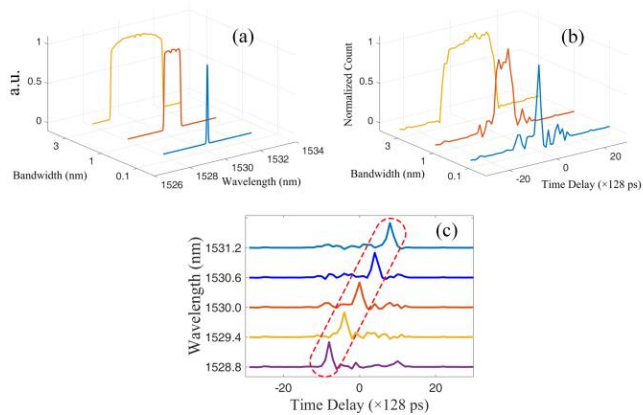

**Fig. 4.** Experiment results when the bandwidth and center-wavelength of BVTF are changed. (a) BVTF's transmission spectra with different filter bandwidths and (b) the corresponding results of the thermal temporal GI; (c) results under different center-wavelengths.

Then, we change the bandwidth and center-wavelength of BVTF to demonstrate the feasibility of the thermal temporal GI scheme. Fig. 4(b) is the results when the center-wavelength of the BVTF is fixed at 1530 nm and its bandwidth is set as 0.1 nm, 1 nm and 3 nm, respectively. The measured transmission spectra of the BVTF is shown in Fig. 4(a) for comparison. It can be seen that the peak of the image broadens with the increasing bandwidth of BVTF. Fig. 4(c) shows the results when the bandwidth is 0.1 nm and the center-wavelength varies from 1528.8 nm to 1531.2 nm with an interval of 0.6 nm. It also can be seen that the position of the peak moves in the time domain with the variation of the center-wavelength of the BVTF. Since photons with shorter wavelength have larger group velocity in the SMF, the peak moves from the left to the right with the increasing center-wavelength. According to Eq. (9), if the center-wavelength changes 0.6 nm, the peak would move ~480 ps (the dispersion of the 50 km SMF is estimated as 800 ps/nm in the calculation). The experimental result shows the average movement is 544 ps. The deviation is due to the difference between the dispersion of the SMF used in the experiment and the value in the estimation. These experimental results show that the profile of transmission spectrum of BVTF can be imaged successfully by the coincidence measurement between the Alice and Bob, demonstrating the feasibility of the thermal temporal GI. Furthermore, the filter BVTF can be equivalently replaced by a grating and a transmitted (or reflected) object. Thus, Bob can recover the image of the object at Alice side by making the correlation of the thermal photons experiencing the spatial dispersion and the temporal dispersion. The resolution of the temporal GI is jointly determined by the timing jitters of the detecting system and the resolution of the grating [15], or the bandwidth of the BVTF in this experiment. In the experiment, the resolution of the current GI is ~140 ps in the time domain. On the other hand, the bandwidth of the thermal source is ~3 nm, leading to a temporal extension of ~2.4 ns, which limits the pixel number in each measurement. We notice that this problem could be solved by generating a broadband single-mode thermal state without any spectral filter [23].

In summary, we have proposed and experimentally demonstrated a scheme of temporal thermal GI over long-distance fibers in proof of principle. The theoretical analysis shows that the mutual coherence function of thermal photons contains the information of Alice's spectral modulation when a large temporal dispersion is introduced at Bob side. As expected, the tailored peak in the coincidence measurement can image the spectral modulation of the photons at Alice side when the bandwidth or center-wavelength of the filter changes. In the experiment, a bandwidth-variable tunable filter is used as an object, which records its spectral information on the spectrum of thermal photons. The experimental results show that the thermal temporal GI can be realized over long-distance fiber links. It can be expected that this scheme also can be used in the imaging of spatial structure of the object, if a spatial dispersion component is introduced at Alice side as Fig. 1.

**Funding** National Key R&D Program of China under Contracts No. 2017 YFA0303700; the National Natural Science Foundation of China under Contracts No. 61575102, No. 91121022, and No. 61621064; the Tsinghua University Initiative Scientific Research Program; the Science and Technology Commission of Shanghai Municipality (Grant No. 16JC1400402)

**References**
1. B. I. Erkmen, and J. H. Shapiro, Adv. Opt. Photon. **2**, 405 (2010).
2. T. B. Pittman, Y. H. Shih, D. V. Strekalov, and A. V. Sergienko, Phys. Rev. A **52**, R3429 (1995).
3. A. F. Abouraddy, B. E. A. Saleh, A. V. Sergienko, and M. C. Teich, Phys. Rev. Lett. **87**, 123602 (2001).
4. R. S. Bennink, S. J. Bentley, and R. W. Boyd, Phys. Rev. Lett. **89**, 113601 (2002).
5. A. Gatti, E. Brambilla, and L. A. Lugiato, Phys. Rev. Lett. **90**, 133603 (2003).
6. A. F. Abouraddy, P. R. Stone, A. V. Sergienko, B. E. A. Saleh, and M. C. Teich, Phys. Rev. Lett. **93**, 213903 (2004).
7. Y. Cai, and S. Zhu, Opt. Lett. **29**, 2716 (2004).
8. A. Valencia, G. Scarcelli, M. D'Angelo, and Y. H. Shih, Phys. Rev. Lett. **94**, 063601 (2005).
9. D. Zhang, Y. H. Zhai, L. A. Wu, and X. H. Chen, Opt. Lett. **30**, 2354 (2005).
10. R. Meyers, K. S. Deacon, and Y. H. Shih, Phys. Rev. A **77**, 041801 (2008).
11. O. Katz, Y. Bromberg, and Y. Silberberg, Appl. Phys. Lett. **95**, 131110 (2009).
12. Y. Bromberg, O. Katz, and Y. Silberberg, Phys. Rev. A **79**, 053840 (2009).
13. F. Ferri, D. Magatti, L. A. Lugiato, and A. Gatti, Phys. Rev. Lett. **104**, 253603 (2010).
14. P. Ryczkowski, M. Barbier, A. T. Friberg, J. M. Dudley, and G. Genty, Nat. Photonics **10**, 167 (2016).
15. S. Dong, W. Zhang, Y. D. Huang, and J. D. Peng, Sci. Rep. **6**, 26022 (2016).
16. Y. H. Shih, *An introduction to quantum optics: photons and biphoton physics* (CRC Press, 2011).
17. G. P. Agrawal, *Nonlinear Fiber Optics*, 5th Ed. (Academic Press, 2013)
18. A. Jeffrey, and D. Zwillinger, *Table of Integrals, Series, and Products*, 6th Ed. (Academic Press, 2003).
19. N. Bruno, A. Martin, T. Guerreiro, B. Sanguinetti, and R. T. Thew, Opt. Express **22**, 017246 (2014).
20. B. Srivathsan, G. K. Gulati, B. Chng, G. Maslennikov, D. Matsukevich, and C. Kurtsiefer, Phys. Rev. Lett. **111**, 123602 (2013).
21. A. Eckstein, A. Christ, P. J. Mosley, and C. Silberhorn, Phys. Rev. Lett. **106**, 013603 (2011).
22. I. Ali-Khan, C. J. Broadbent, and J. C. Howell, Phys. Rev. Lett. **98**, 060503 (2007).
23. P. J. Mosley, J. S. Lundeen, B. J. Smith, P. Wasylczyk, A. B. U'Ren, C. Silberhorn, and I. A. Walmsley, Phys. Rev. Lett. **100**, 133601 (2008).

## References

1. B. I. Erkmen, and J. H. Shapiro, Ghost imaging: from quantum to classical to computational, Adv. Opt. Photon. **2**, 405-450 (2010).
2. T. B. Pittman, Y. H. Shih, D. V. Strekalov, and A. V. Sergienko, Optical imaging by means of two-photon quantum entanglement, Phys. Rev. A **52**, R3429-R3432 (1995).
3. A. F. Abouraddy, B. E. A. Saleh, A. V. Sergienko, and M. C. Teich, Role of entanglement in two-photon imaging, Phys. Rev. Lett. **87**, 123602 (2001).
4. R. S. Bennink, S. J. Bentley, and R. W. Boyd, "Two-photon" coincidence imaging with a classical source, Phys. Rev. Lett. **89**, 113601 (2002).
5. A. Gatti, E. Brambilla, and L. A. Lugiato, Entangled imaging and wave-particle duality: from the microscopic to the macroscopic realm, Phys. Rev. Lett. **90**, 133603 (2003).
6. A. F. Abouraddy, P. R. Stone, A. V. Sergienko, B. E. A. Saleh, and M. C. Teich, Entangled-photon imaging of a pure phase object, Phys. Rev. Lett. **93**, 213903 (2004).
7. Y. Cai, and S. Zhu, Ghost interference with partially coherent radiation, Opt. Lett. **29**, 2716-2718 (2004).
8. A. Valencia, G. Scarcelli, M. D'Angelo, and Y. H. Shih, Two-photon imaging with thermal light, Phys. Rev. Lett. **94**,063601 (2005).
9. D. Zhang, Y. H. Zhai, L. A. Wu, and X. H. Chen, Correlated two-photon imaging with true thermal light, Opt. Lett. **30**, 2354-2356 (2005).
10. R. Meyers, K. S. Deacon, and Y. H. Shih, Ghost-imaging experiment by measuring reflected photons, Phys. Rev. A **77**, 041801 (2008).
11. O. Katz, Y. Bromberg, and Y. Silberberg, Compressive ghost imaging, Appl. Phys. Lett. **95**, 131110 (2009).
12. Y. Bromberg, O. Katz, and Y. Silberberg, Ghost imaging with a single detector, Phys. Rev. A **79**, 053840 (2009).
13. F. Ferri, D. Magatti, L. A. Lugiato, and A. Gatti, Differential Ghost Imaging, Phys. Rev. Lett. **104**, 253603 (2010).
14. P. Ryczkowski, M. Barbier, A. T. Friberg, J. M. Dudley, and G. Genty, Ghost imaging in the time domain, Nat. Photonics **10**, 167-170 (2016).
15. S. Dong, W. Zhang, Y. D. Huang, and J. D. Peng, Long-distance temporal quantum ghost imaging over optical fibers, Sci. Rep. **6**, 26022 (2016).
16. Y. H. Shih, An introduction to quantum optics: photons and biphoton physics (CRC Press, 2011).
17. G. P. Agrawal, Nonlinear Fiber Optics, 5th Ed. (Academic Press, 2013)
18. A. Jeffrey, and D. Zwillinger, Table of Integrals, Series, and Products, 6th Ed. (Academic Press, 2003).
19. N. Bruno, A. Martin, T. Guerreiro, B. Sanguinetti, and R. T. Thew, Pulsed source of spectrally uncorrelated and indistinguishable photons at telecom wavelengths, Opt. Express **22**, 17246-17253 (2014).
20. B. Srivathsan, G. K. Gulati, B. Chng, G. Maslennikov, D. Matsukevich, and C. Kurtsiefer, Phys. Rev. Lett. **111**, 123602 (2013).
21. A. Eckstein, A. Christ, P. J. Mosley, and C. Silberhorn, Narrow band source of transform-limited photon pairs via four-wave mixing in a cold atomic ensemble, Phys. Rev. Lett. **106**, 013603 (2011).
22. I. Ali-Khan, C. J. Broadbent, and J. C. Howell, Large-alphabet quantum key distribution using energy-time entangled bipartite states, Phys. Rev. Lett. **98**, 060503 (2007).
23. P. J. Mosley, J. S. Lundeen, B. J. Smith, P. Wasylczyk, A. B. U'Ren, C. Silberhorn, and I. A. Walmsley, Heralded generation of ultrafast single photons in pure quantum states, Phys. Rev. Lett. **100**, 133601 (2008).